\documentclass[aps,prl,twocolumn,superscriptaddress]{revtex4-1}
\usepackage{graphics}
\usepackage{graphicx}
\usepackage{color}
\usepackage{epstopdf}
\usepackage{amsmath}
\usepackage{cancel}
\begin{document}
\title{Elastocapillary bending of microfibers around liquid droplets} 
\author{Rafael D. Schulman}
\affiliation{Department of Physics and Astronomy, McMaster University, 1280 Main St. W, Hamilton, ON, L8S 4M1, Canada.}
\author{Amir Porat}
\affiliation{Laboratoire de Physico-Chimie Th\'eorique, UMR CNRS Gulliver 7083, ESPCI Paris, PSL Research University, 75005 Paris, France.}
\author{Kathleen Charlesworth}
\affiliation{Department of Physics and Astronomy, McMaster University, 1280 Main St. W, Hamilton, ON, L8S 4M1, Canada.}
\author{Adam Fortais}
\affiliation{Department of Physics and Astronomy, McMaster University, 1280 Main St. W, Hamilton, ON, L8S 4M1, Canada.}
\author{Thomas Salez}
\affiliation{Laboratoire de Physico-Chimie Th\'eorique, UMR CNRS Gulliver 7083, ESPCI Paris, PSL Research University, 75005 Paris, France.}
\affiliation{Global Station for Soft Matter, Global Institution for Collaborative Research and Education, Hokkaido University, Sapporo, Hokkaido 060-0808, Japan.}
\author{Elie Rapha\"{e}l}
\affiliation{Laboratoire de Physico-Chimie Th\'eorique, UMR CNRS Gulliver 7083, ESPCI Paris, PSL Research University, 75005 Paris, France.}
\author{Kari Dalnoki-Veress}
\email{dalnoki@mcmaster.ca}
\affiliation{Department of Physics and Astronomy, McMaster University, 1280 Main St. W, Hamilton, ON, L8S 4M1, Canada.}
\affiliation{Laboratoire de Physico-Chimie Th\'eorique, UMR CNRS Gulliver 7083, ESPCI Paris, PSL Research University, 75005 Paris, France.}
\date{\today}
\begin{abstract}
We report on the elastocapillary deformation of flexible microfibers in contact with liquid droplets. A fiber is observed to bend more as the size of the contacting droplet is increased. At a critical droplet size, proportional to the bending elastocapillary length, the fiber is seen to spontaneously wind around the droplet. To rationalize these observations, we invoke a minimal model based on elastic beam theory, and find agreement with experimental data. Further energetic considerations provide a consistent prediction for the winding criterion. 
\end{abstract}
\maketitle

Wetting of liquids on fibrous materials is central to a wide variety of natural and industrial phenomena such as the coalescence of wet hairs~\cite{Bico2004, Py2007}, the drying of textiles~\cite{Sauret2015}, the altered mechanical properties of dewy spider silk~\cite{Vollrath1989,Elettro2015,Elettro2016}, the defense mechanism of a species of beetle~\cite{Eisner2000}, and the bundling of carbon nanotubes and nanowires during processing~\cite{Chakrapani2003,Lau2003,Dev2007,Pokroy2009}. In some of these examples, the fibers are sufficiently flexible that capillary forces induce large-scale deformations -- a phenomenon also observed in other geometries such as a drop contacting a flexible solid strip~\cite{Rivetti2012}. The bending elastocapillary length $L_{\textrm{BC}} = \sqrt{E\,r^3/\gamma}$ is the natural length scale that emerges when balancing elastic bending and capillarity, where $E$ is the Young's modulus of the fiber, $r$ is the fiber radius, and $\gamma$ is the liquid-air surface tension~\cite{Roman2010,Liu2012}. A slender structure is significantly deformed by capillary forces if the length scale over which these forces act is larger than $L_{\textrm{BC}}$~\cite{Roman2010}. To understand the wetting of fibers, several model experiments have been carried out, focusing on droplets between slender flexible structures, where material stiffness and geometry dictate the final wetting configuration~\cite{Bico2004,Bedarkar2010,Duprat2012,Protiere2012,Duprat2015}.

Despite its simplicity, even the problem of a single droplet atop an undeformable cylinder is interesting as there are two possible equilibrium states: an axisymmetric ``barrel'' configuration and a non-axisymmetric ``clam-shell"~\cite{Carroll1976,Carroll1984,McHale1997,McHale2002,Chou2011}. It is then not surprising that the case of a flexible fiber interacting with a liquid is a rich subject of study, showcasing complexity and stunning examples of self-assembly~\cite{Roman2010,marchand2012,Evans2013,Fargette2014,Elettro2015,Elettro2016}. In a series of beautiful experiments, droplets were placed on taut elastomeric fibers, and reached the barrel configuration~\cite{Elettro2015,Elettro2016}. With reduced tension, capillary forces cause the fiber to buckle inside the droplet if the radius of the latter exceeds roughly $L_{\textrm{BC}}$. As the fiber is slackened, it coils inside the droplet which acts as a windlass to maintain tension. However, for a smaller droplet-to-fiber radius ratio, or for less-wettable conditions, the clam-shell configuration may be more favourable than the barrel~\cite{Chou2011}. In such a case, a soft fiber may instead wind \textit{around} the surface of a droplet without experiencing a buckling transition~\cite{Roman2010}. As argued by Roman and Bico, the reduction in surface energy upon winding exceeds the bending penalty if the droplet radius is larger than $\sim L_{\textrm{BC}}$. This is reminiscent of DNA molecules wrapping around histone octamers to form compact structures within the nucleus~\cite{Marky1991, Roman2010}. When a droplet is wound by a fiber, it changes from a spherical shape to a lens configuration in which the fiber is positioned at the equator. Roman and Bico focused on understanding the lenticular geometry but did not experimentally test the winding criterion. Furthermore, there has been no experimental work investigating the deformation below the winding threshold.

In this Letter, we study the elastocapillary bending of microfibers induced through contact with liquid droplets. By gradually increasing the droplet radius, experiments reveal that fibers become increasingly bent around the droplet before the winding criterion is met. We invoke a minimal model based on elastic beam theory to quantitatively understand these observations. Using this model, we estimate the winding threshold, and find it to be in agreement with data as well as a prediction from simple energetic considerations. Finally, for adhesive polymer fibers in the wound state, removal of the droplets leaves behind self-assembled dry polymer microcoils.

The fibers were made using two different materials: Polystyrene (PS) with molecular weight $M_n=25$~kg/mol (Polymer Source Inc.), which is a glass at room temperature; and Styrene-Isoprene-Styrene (SIS) triblock copolymer (14\,\% styrene content, Sigma-Aldrich), which is a physically crosslinked elastomer at room temperature. PS fibers were made by dipping a micropipette into a PS melt held at 170~$^\circ$C and then quickly pulling the pipette out, resulting in fibers with radii $2\mathrm{\ \mu m}<r<6\mathrm{\ \mu m}$, as measured with optical microscopy. A similar procedure was used to produce SIS fibers. However, the fibers were instead pulled from a concentrated solution of SIS and toluene. The SIS fibers were prepared with $5\mathrm{\ \mu m}<r<25\mathrm{\ \mu m}$. Fiber radii were uniform to within 10\% over the length used.  The Young's modulus of PS is found in the literature to be 3.4~GPa~\cite{Brandrup1999}, and the Young's modulus of SIS was determined to be $0.80 \pm 0.15$~MPa by pulling on fibers and measuring the resultant forces using a micropipette deflection technique~\cite{Colbert2009}.

\begin{figure}
\includegraphics[width=1 \columnwidth]{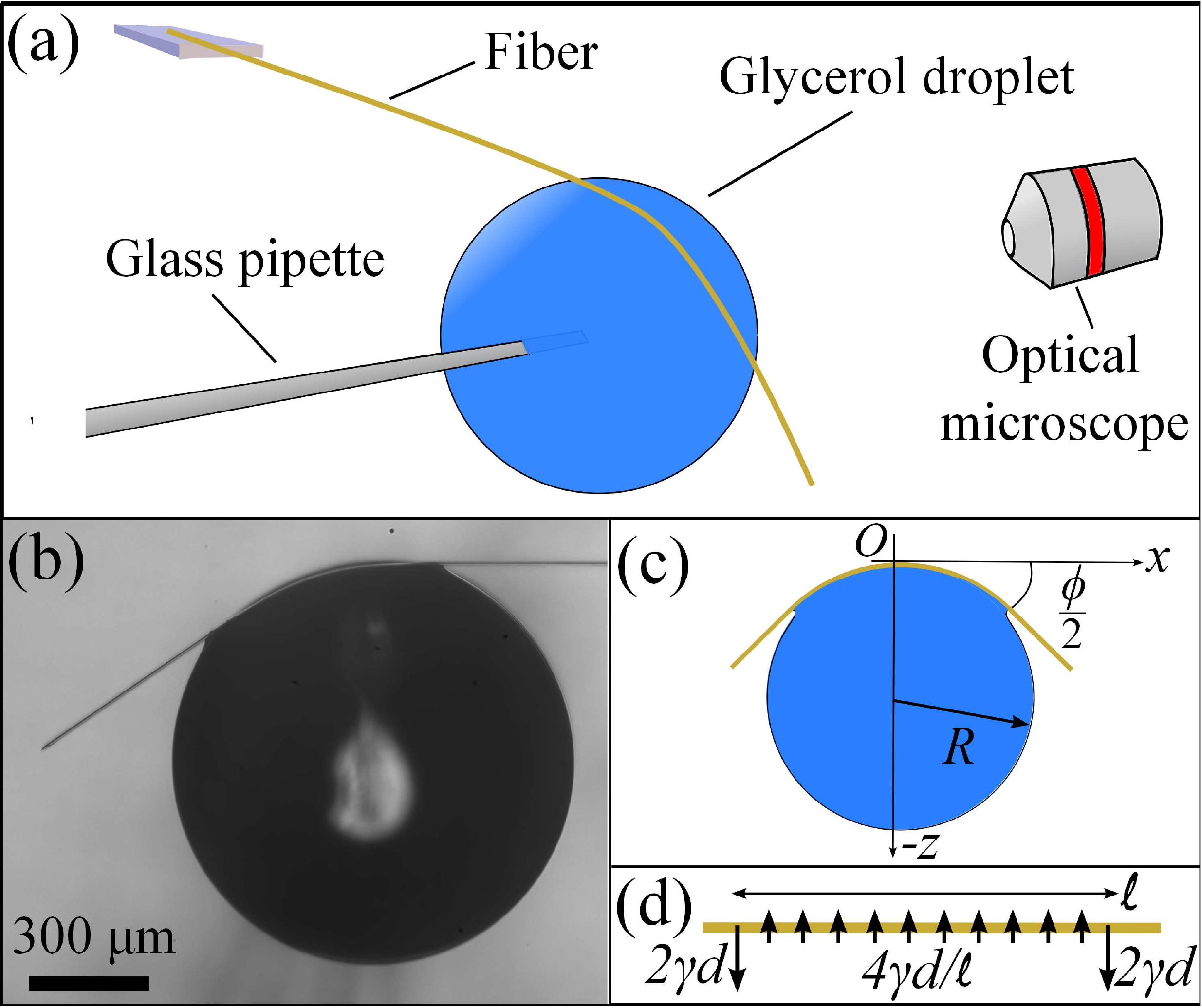}
\caption{(a) Schematic of the experiment for PS fibers. (b) Optical image. (c) Schematic of the tilted optical image with notations. (d) Idealized force distribution within the contact region, as described in text.}
\label{fig1}
\end{figure}

The experimental setup for the PS fibers is depicted in Fig.~\ref{fig1}(a). Initially, both ends of a given fiber are taped to two separate silicon pieces, which are pulled apart to hold the fiber taut. A glass micropipette is used to support a glycerol droplet, whose surface tension with air is $\gamma = 63$~mN/m~\cite{korosi1981,lide2004}. The pipette is connected to a syringe filled with glycerol, which allows us to precisely control the size of the droplet. The fiber is then brought into contact with the liquid droplet. Once in contact, the fiber is snipped at one end so that it is free to move and no longer under tension. The system is imaged with an optical microscope along the pipette axis and perpendicularly to the fiber. Note that we work much below the capillary length, such that gravitational effects can be safely ignored~\cite{Roman2010}.
A typical optical image is shown in Fig.~\ref{fig1}(b), where it is clear that the fiber deforms through its interaction with the droplet. We can analyze these images to extract the angle $\phi$ through which the fiber is bent, the droplet radius $R$  -- marginally modified by the fiber since $r\ll R$ -- and the shape of the fiber in the contact region, shown schematically in Fig.~\ref{fig1}(c). Outside the region of liquid contact, the fiber is straight.   

\begin{figure}
\includegraphics[width=1 \columnwidth]{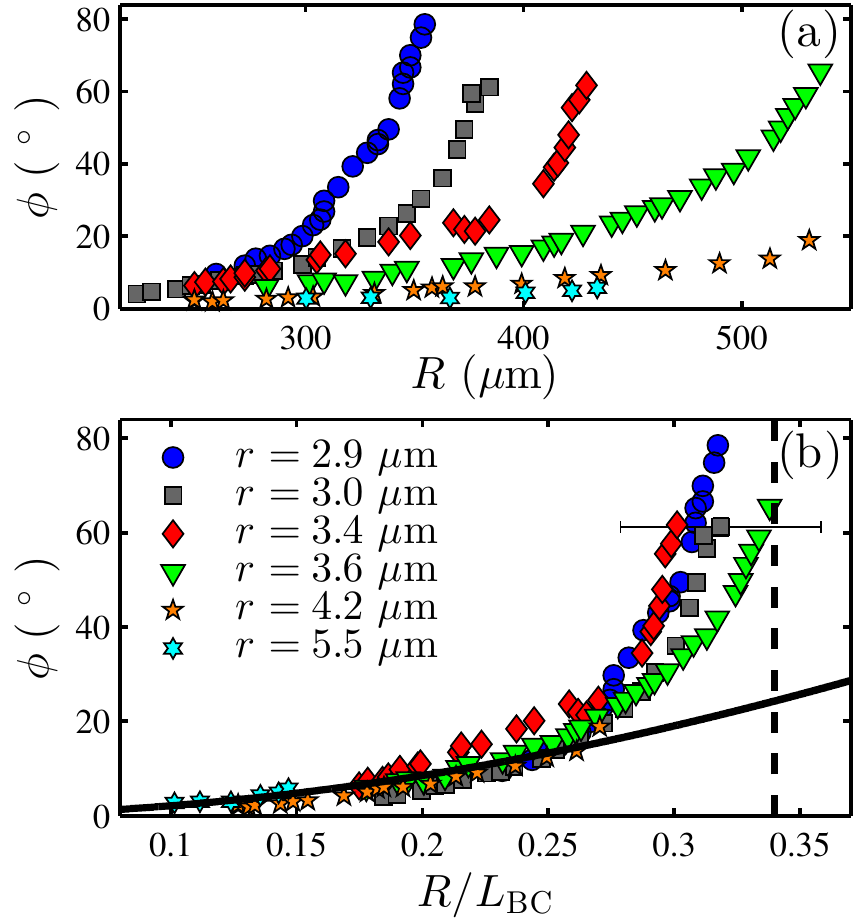}
\caption{Deformation angle (Fig.~\ref{fig1}(c)) of six PS fibers, of radii $r$ as indicated, as a function of: (a) the droplet radius; and (b) the droplet radius normalized by the bending elastocapillary length. The vertical error bars are comparable to the marker size. One representative horizontal error bar is shown. The uncertainty in $R/L_{\textrm{BC}}$ is dominated by the error in the measurement of $r$, and would thus be in the same direction for all the data of a given fiber. The solid curve is the best fit to Eq.~(\ref{tan}) to small-$\phi$ data where the scaling
$l\propto R$ is valid [34]. The dashed line corresponds to Eq.~(\ref{Rf}) with $R_{\textrm{f}}=R$, using the fit parameters ($\mu,\lambda$) of Figs.~\ref{fig2}(b) and~\ref{fig3}.}
\label{fig2}
\end{figure} 
We begin the experiments with a relatively small droplet ($R \sim 250\ \mathrm{\mu}$m), and observe that there is only a minor deformation of the fiber. By applying pressure through the syringe the droplet grows, and as a result, the fiber increases its contact with the droplet, and the deformation angle $\phi$ increases. The data are shown in Fig.~\ref{fig2}(a), in which $\phi$ is plotted as a function of $R$ for different fiber radii $r$. We observe that for a given droplet radius, thicker fibers exhibit smaller deformations, due to the higher bending moduli. Since the bending of the fiber is caused by the interaction with a liquid droplet, capillarity is the driving mechanism. $L_{\textrm{BC}}$ is thus expected to be the relevant length scale of the problem, as confirmed by the collapse of the data (within error) in Fig.~\ref{fig2}(b) where $R$ is normalized by $L_{\textrm{BC}}$. 

Although the problem of a liquid droplet deforming an elastic beam has been solved analytically in 2D~\cite{Neukirch2013}, the 3D analogue is considerably more complicated. Therefore, we focus on the essential physical ingredients only and propose the following minimal model. We consider an idealized force distribution acting on the elastic fiber, in the contact region of length $\ell$, as shown in Fig.~\ref{fig1}(d). As seen in Fig.~\ref{fig1}(b), at each of the two edges of the contact region, there is a meniscus force pulling the beam inwards towards the droplet. Due to its capillary origin, this force is expected to scale as $2\gamma d$, where the factor 2 accounts for the two sides of one meniscus, and where $d$ is a length scale characterizing the lateral extent of the meniscus -- while incorporating as well an unknown dimensionless geometrical prefactor of order unity. As $d$ is expected to be substantially smaller than $R$, we describe the meniscus force as being point-like. To maintain a zero net force at equilibrium, there must also be a force in the contact region pushing the beam outwards. For simplicity, we assume it to be uniformly distributed with the linear density $f=4\gamma d/\ell$ whose integral over the contact region balances the two meniscus forces. We may now solve for the fiber profile $z(x)$ in the contact region, invoking the small-deformation bending-beam equation~\cite{timoshenko1951}: $Bz''''=f$, where the prime denotes the derivative with respect to $x$, and where $B=\pi Er^4/4$ is the bending modulus of the fiber. Using the no-torque boundary condition $z''(\ell/2)=0$, the origin definition $z(0)=0$, and the even symmetry $z(x)=z(-x)$, the solution reads:
\begin{equation}
z(x)=\frac{d\ell^3}{\pi rL_{\textrm{BC}}^{\,2}}\left[\frac23\left(\frac{x}{\ell}\right)^4-\left(\frac{x}{\ell}\right)^2\right]\ .
\end{equation}
From this solution, it is straightforward to determine the outer slope $\tan(\phi/2)=-z'(\ell/2)$, and the central radius of curvature $R_{\textrm{f}}=-1/z''(0)$, that characterize the deformation. Motivated by experimental data in the small-$\phi$ limit~\cite{SI}, we further assume the scale-separation relations: $\ell=\lambda R$, and $d=\mu r$, where $\lambda$ and $\mu$ are dimensionless constants. In such a description, one finally gets:
\begin{eqnarray}
\label{tan}
\mathrm{tan}\left(\frac{\phi}{2}\right) &=& \frac{2\mu\lambda^2}{3\pi}\left(\frac{R}{L_{\textrm{BC}}}\right)^2\\
\frac{R_{\textrm{f}}}{R} &=&  \frac{\pi}{2\mu\lambda}\left(\frac{ L_{\textrm{BC}}}{R}\right)^2\ .
\label{Rf}
\end{eqnarray}

\begin{figure}
\includegraphics[width=1 \columnwidth]{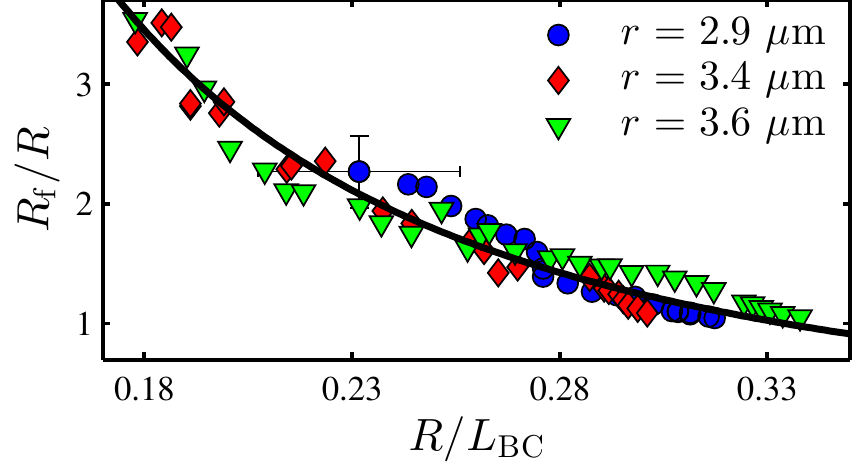}
\caption{Central radius of curvature of the fiber normalized by the droplet radius, as a function of the droplet radius normalized by the bending elastocapillary length. One representative set of error bars is shown. The vertical error is dominated by the determination of $R_{\textrm{f}}$. The horizontal error bars are the same as in Fig.~\ref{fig2}. The solid curve corresponds to the best fit to Eq.~(\ref{Rf}).}
\label{fig3}
\end{figure}

From Eq.~(\ref{tan}), we confirm that $\phi$ is a function of $R/L_{\textrm{BC}}$ only. Furthermore, the best fit of Eq.~(\ref{tan}) to small-angle data ($\phi \lesssim 15^\circ$ and $R/L_{\textrm{BC}}<0.25$ where the scaling $l \propto R$ is valid~\cite{SI}) is shown in Fig.~\ref{fig2}(b). We see that the fit describes the data well for small $\phi$ but fails to capture the large-angle data. This is to be expected since a small-deformation theory was employed and the contact region $l$ deviates from the assumed scaling at large deformation. Similarily, Eq.~(\ref{Rf}) can be tested by measuring $R_{\textrm{f}}$, the central radius of curvature of the fiber, for various experiments~\cite{SI}. As shown in Fig.~\ref{fig3}, the collapse of the data for different fiber thicknesses is consistent with the prediction, and the best fit to Eq.~(\ref{Rf}) is excellent, even at large angles. We assume the success at large angles is because $R_{\textrm{f}}$ is a local quantity in the central region where $z'$ is small. Moreover, from the fit parameters, we find $\lambda\approx0.7$ and $\mu\approx21$ for the two unknown prefactors characterizing the contact and meniscus lateral extents for a PS fiber. These values imply that the contact region is comparable to the droplet size, while the meniscus is roughly an order of magnitude larger than the fiber diameter. These results are consistent with what we would expect from the optical images~\cite{SI}.  Finally, we note that if the curvature of the fiber matches that of the droplet, this corresponds to the fiber winding around the droplet. Thus, winding should occur when $R_\textrm{f} \rightarrow R$ which, using Eq.~(\ref{Rf}) and the values of $\mu$ and $\lambda$, corresponds to $R/L_{\textrm{BC}} = 0.34 \pm 0.02$ for PS. As indicated by the vertical dashed line in Fig.~\ref{fig2}(b), this prediction is consistent with the data, as the winding angle appears to diverge at this point. 

Eq.~(\ref{Rf}) suggests that there is  a critical droplet size for which fibers will completely wind around the droplet. Indeed, we verify experimentally that when a fiber is placed into contact with a sufficiently large droplet, the fiber does  spontaneously wind as previously found by Roman and Bico~\cite{Roman2010}. Roman and Bico showed that this transition can be simply explained from energetic considerations. Upon winding, the free energy of the system can be written per unit length, with only two terms~\cite{SI}: the surface energy of the system, which is negative (i.e. reduced compared to the unwound state) due to contact between the droplet and the fiber and scales like $-\gamma r$, and an energetic penalty associated with the bending of the fiber, which scales like $B/2R^2$. Winding occurs when it lowers the free energy of the system, which results in the winding criterion
$R > \alpha  L_{\textrm{BC}}$, where $\alpha$ depends on the details of the wetting geometry~\cite{SI}. In the limit $r \ll R$, the microscopic wetting picture is equivalent to that of a cylinder on the surface of a liquid bath, where the liquid surface is flat and Young's law is satisfied~\cite{SI}. Considering this, we find $\alpha = \sqrt{\pi / 16 [\mathrm{sin}\theta_y + (\pi - \theta_y)\mathrm{cos}\theta_y]}$, where $\theta_y$ is the Young's angle of the liquid on the solid. Inserting the measured values of $\theta_y$, we finally evaluate $\alpha_\mathrm{PS} = 0.37 \pm 0.01$ and
$\alpha_\mathrm{SIS} = 0.40 \pm 0.01$ for PS and SIS respectively. The predicted value for PS compares closely with the value of $0.34 \pm 0.02$ attained from Eq.~(\ref{Rf}). 

\begin{figure}[t]
\includegraphics[width=1 \columnwidth]{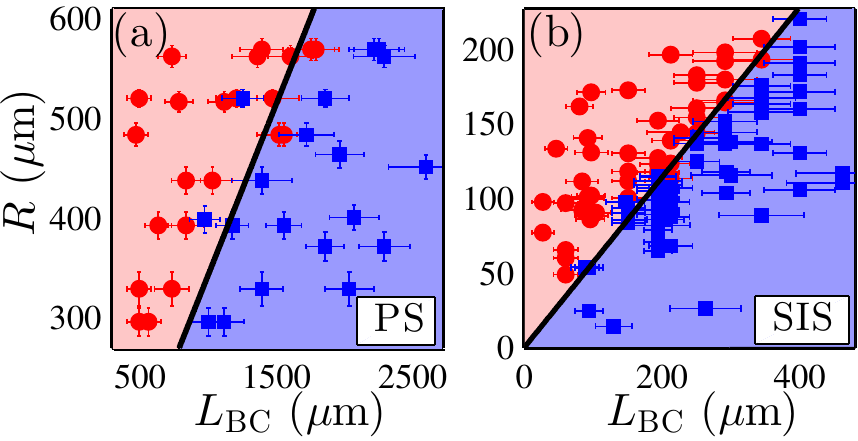}
\caption{The winding phase diagram for (a) PS and (b) SIS fibers. Circle markers denote winding events and square markers indicate experiments where no winding occurred. The phase boundary (solid black line) is fit to a straight line passing through the origin. }
\label{fig4}
\end{figure} 

To test the predicted winding threshold for PS, we utilize the same experimental design as depicted in Fig.~\ref{fig1}(a), and observe whether or not the fiber spontaneously winds around the droplet. This experiment allows us to construct a winding phase diagram of $R$ as a function of $L_{\textrm{BC}}$ as shown in Fig.~\ref{fig4}(a). The circular data points correspond to winding events and square data points denote experiments in which no winding occurred. The black line is the best fit to a phase boundary represented by a line passing through the origin. From the fit, we extract $\alpha_\mathrm{PS} = 0.34 \pm 0.04$, which is in excellent agreement with the energetically derived value of $0.37 \pm 0.01$ and the value of $0.34 \pm 0.02$ attained from Eq.~(\ref{Rf}).

\begin{figure}[t]
\includegraphics[width=1 \columnwidth]{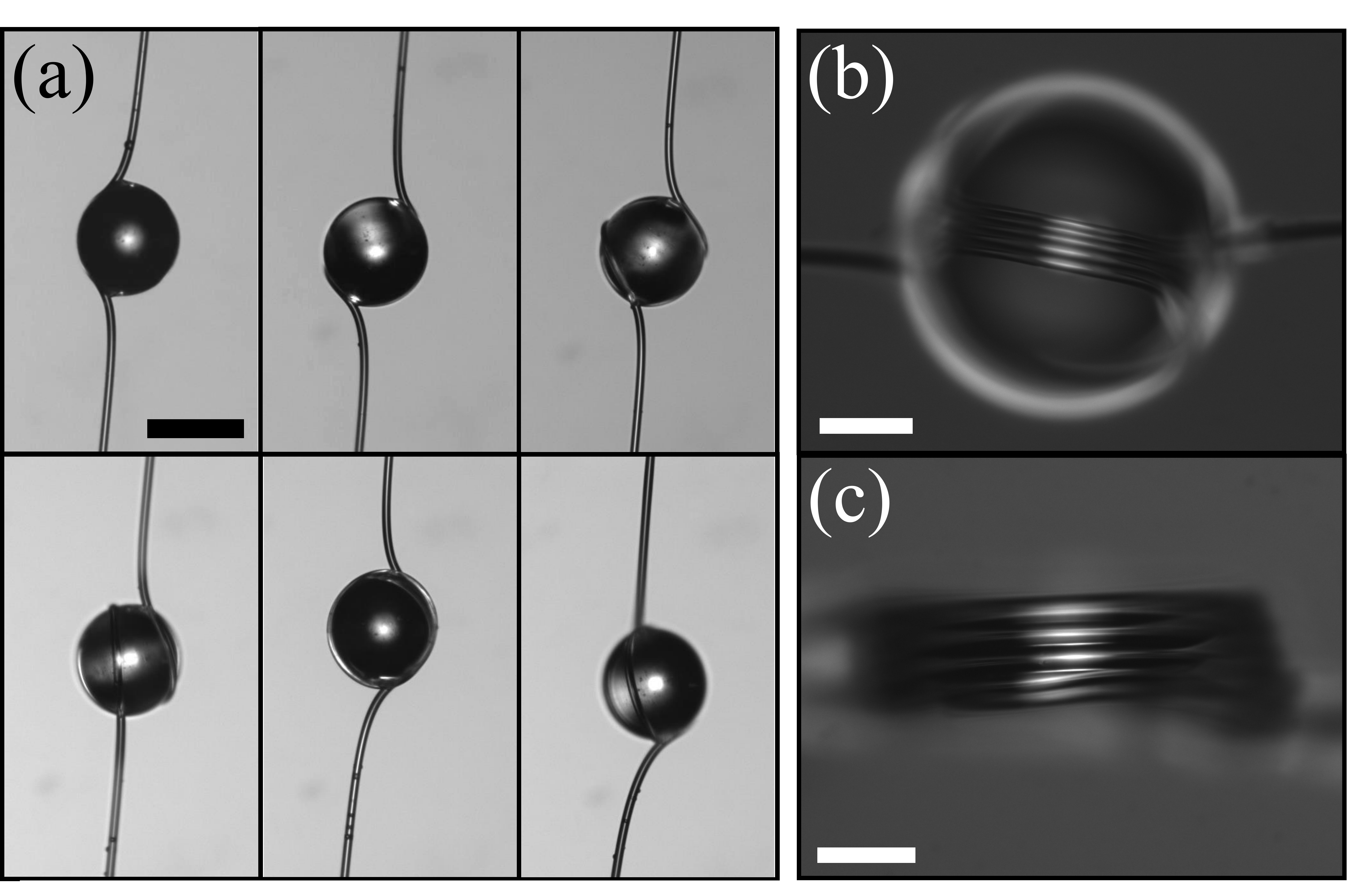}
\caption{(a) Image sequence of a pendant droplet being wound by an SIS fiber (to be read: top row, left to right, bottom row, left to right). In the last two frames, the fiber is wound once around the droplet but the system is rotated to provide different perspectives. Scale bar = 200 $\mu$m (b) Optical microscopy image of an SIS fiber wound five times around a single droplet. Scale bar = 100 $\mu$m. (c) Same as (b) after removing the droplet.}
\label{fig5}
\end{figure}

Since SIS has a modulus three orders of magnitude smaller than PS, the bending elastocapillary length for the fiber radii used is much smaller than for PS. Due to this, significantly smaller droplets are required for the experiment and it no longer becomes practical to suspend these droplets at the tip of a micropipette. Instead, we keep the SIS fibers taut between two supports and directly transfer a glycerol droplet onto the fiber. In doing so, the droplet assumes a clam-shell configuration~\cite{McHale2002}, but is close to spherical since $R\gg r$. Subsequently, the supports are brought in closer together, and as the fiber slackens, the droplet gets wound by the fiber if the droplet is sufficiently large. The resulting phase diagram is shown in Fig.~\ref{fig4}(b). Once again, the phase boundary is well fit by a line passing through the origin. However, we find $\alpha_\mathrm{SIS} = 0.57 \pm 0.05$, significantly larger than the value predicted using energy considerations: $0.40 \pm 0.01$. The origin of this discrepancy is likely due to the fact that, in this case, the droplets are pendant on the taut fiber which has no free ends for winding. In Fig.~\ref{fig5}(a), we show a sequence of images showcasing how winding occurs in this geometry~\cite{SI}. In this sequence, it is evident that the fiber not only bends when in contact with the droplet (as in the PS case), but retains some curvature beyond the contact patch. This additional bending cost must be included in the energy argument leading to the winding criterion. From the optical images, we note that the integrated curvature outside of contact is approximately equal to to that in contact. Thus, if we argue that the two bending costs are roughly equal in magnitude~\cite{SI}, we arrive at $\alpha_\mathrm{SIS} \sim 0.56$ as a rough estimate, consistent with the experimentally measured value.

In the SIS experiments, as the supports are brought closer together, the fiber continuously winds around the droplet. The resulting fiber loops are packed closely together and produce stunning coils. An example is shown in Fig.~\ref{fig5}(b), where we observe five tightly wound loops of fiber on the surface of a droplet. As the SIS fibers are sticky, the liquid can be removed (by, for instance, dissolving away the liquid in a volatile solvent) while still leaving the fiber structure intact, as shown in Fig.~\ref{fig5}(c). 

We have investigated the elastocapillary interaction between liquid droplets and thin flexible fibers. The fibers develop a contact region with the liquid and become deformed by capillary forces. We quantify the resultant deformation and find the ratio of the droplet size to the bending elastocapillary length to be the relevant variable. To gain further insight into the fiber bending, we present a minimal model based on elastic-beam theory, which captures the scalings for the deformation angle and central curvature of the bent fiber. Furthermore, fibers are seen to spontaneously wind around droplets roughly larger than the bending elastocapillary length. The winding criterion is correctly predicted using the beam theory model  but also independently from simple energetic considerations.
\newline

The authors thank Benoit Roman and Jos\'e Bico for interesting discussions. They acknowledge financial support from the Natural Science and Engineering Research Council of Canada, the \'Ecole Normale Sup\'erieure of Paris, the ESPCI Joliot Chair, the LabEx ENS-ICFP: 
ANR-10-LABX-0010/ANR-10-IDEX-0001-02 PSL, and the Global Station for Soft Matter -- a project of Global Institution for Collaborative Research and Education at Hokkaido University.


%

\pagebreak
\widetext

\renewcommand{\thefigure}{S\arabic{figure}}
\setcounter{figure}{0}
\renewcommand{\theequation}{S\arabic{equation}}
\setcounter{equation}{0}

\section{Supplemental Information for ``Elastocapillary bending of microfibers around liquid droplets''} 

\bigskip

\section{Analysis of Contact Region and Microfiber Shape}
 
\begin{figure}[h]
     \includegraphics[width=\textwidth]{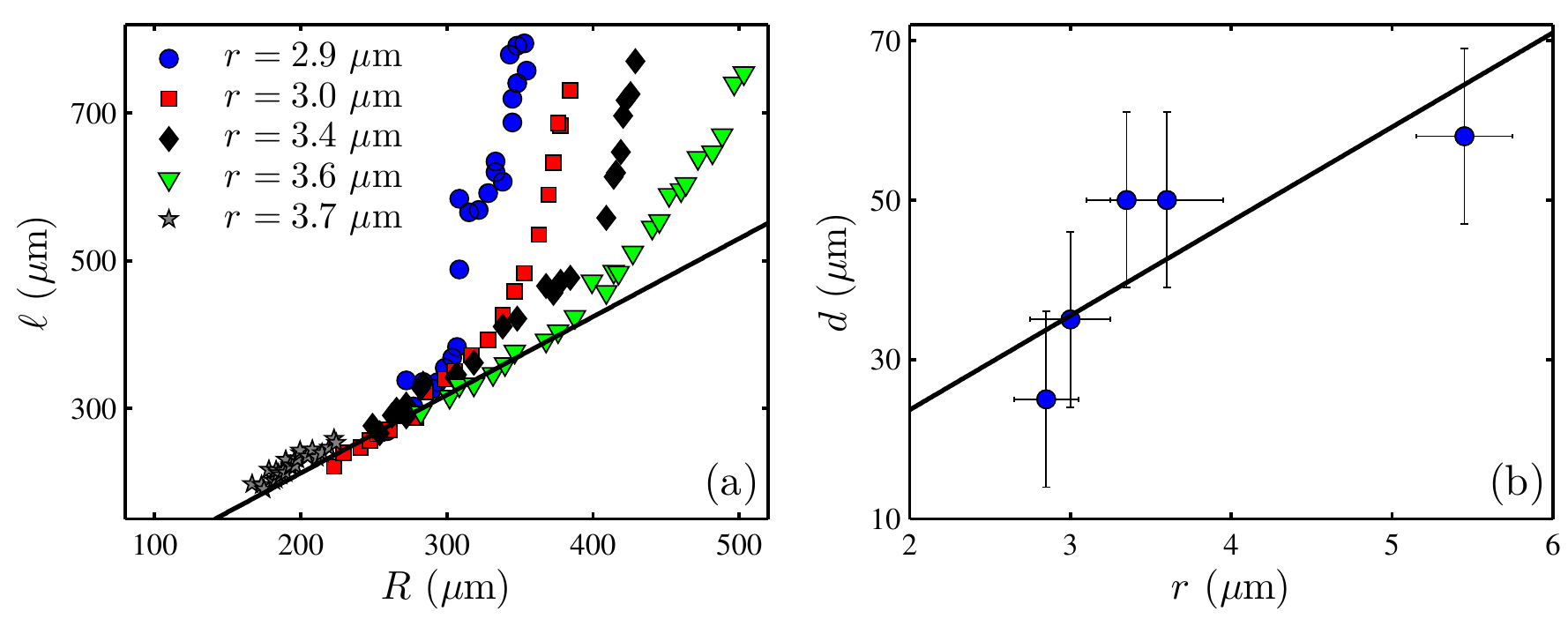}
\caption{(a) The arc length of the wetted contact region between droplet and fiber as a function of droplet radius. The solid line is a straight line passing through the origin to demonstrate that $\ell \propto R$ is valid in the initial regime. (b) The meniscus size as a function of fiber radius.  The solid line is a straight line passing through the origin to demonstrate that $d \propto r$ is consistent with the data.}
\label{figS1}
\end{figure}

To analyze the wetting region between the droplet and the fiber, we begin by thresholding the image to black and white in MATLAB. The bounds of the wetting region are inputted manually, and subsequently, the contour of this region is detected. As such, we can extract the arc length of the contact region $\ell$. For several fiber radii, we plot the data for $\ell$ as a function of droplet radius in Fig.~\ref{figS1}(a). A solid line passing through the origin is drawn alongside the data. The line describes the low-$R$ data very well, demonstrating that the empirical scaling $\ell \propto R$ in the low-$\phi$ limit is valid. The data begins to deviate from this scaling when $\phi \gtrsim 15^\circ$.

To validate our assumption that $d \propto r$, we measure the length of the meniscus region from the images for several fiber radii with $R$ held roughly constant. The results of this analysis are displayed in Fig.~\ref{figS1}(b), which shows that $d$ increases with $r$.  A solid line passing through the origin is drawn to show that the data is consistent with the assumed scaling $d \propto r$. For a given fiber, $d$ is not found to depend on $R$.

\begin{figure}
     \includegraphics[width=6cm]{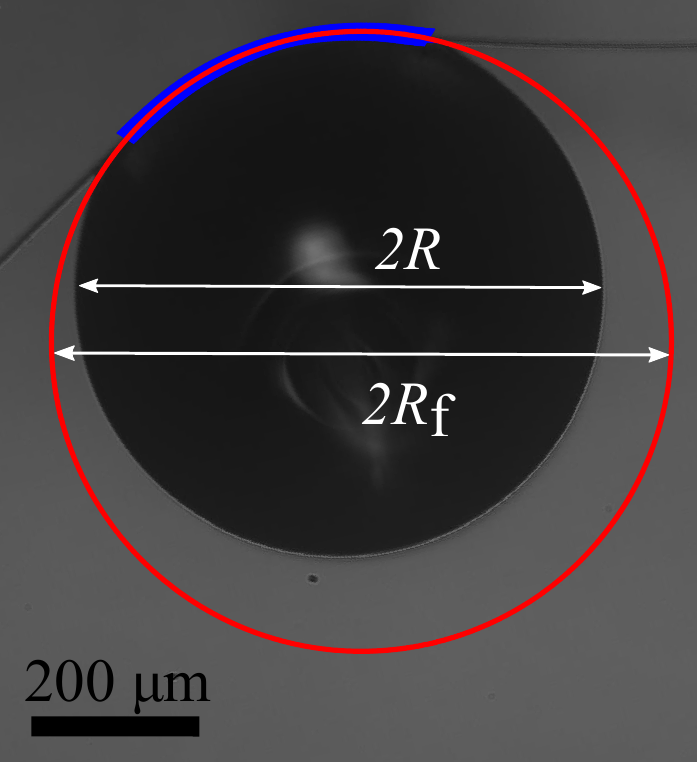}
\caption{A fiber being deformed by a droplet. The contour of the central region of the fiber in contact with the droplet is detected through image analysis (blue) and fit to a circle (red circle).}
\label{figS2}
\end{figure}

We may also fit the central part of this wetting region to a circle to extract $R_\mathrm{f}$. In these fits, we exclude the region nearest to where the fiber exits contact with the liquid, as the fiber in this region is observed to be changing curvature towards becoming straight. A sample fit is shown in Fig.~\ref{figS2}, where the blue region corresponds to the contour detected through image analysis, and the red circle is the fit to that data. We see that the central region of the fiber assumes a curvature of $R_\mathrm{f} >R$.

\section{Energetic Considerations of the Winding Criterion}
As outlined by Roman and Bico~\cite{Roman2010}, the winding threshold can be predicted from simple energetic considerations.  The transition can be explained considering a two-state model where a fiber and a droplet are either in isolation or in the wound state. Upon winding, the surface energy of the system is reduced due to contact between the droplet and the fiber by an amount $-2\gamma\beta r$ per unit length, where $\beta$ is a prefactor which depends on the details of the wetting geometry. Note that since we are in the regime $r<<R$, the droplet remains nearly spherical after being wound, and any change in surface energy due to a global change in shape of the droplet is neglible (as will be demonstrated in the next section). The energetic penalty associated with winding around the droplet is an increase in bending energy of the fiber given by $B/2R^2$ per unit length, where $B = \pi E r^4/4$ is the bending modulus of the fiber. Thus, the total energy change upon winding is:
\begin{equation}
\Delta E = -2\gamma\beta r +\frac{\pi E r^4 }{8R^2} \; .
\label{delta_E}
\end{equation}
Winding occurs when it lowers the free energy of the system, which results in the winding criterion:
\begin{equation}
 R > \alpha  L_\mathrm{BC} \;,
\label{winding}
\end{equation}
where $\alpha = \sqrt{\pi/16\beta}$.  To attain a prediction for $\beta$, we must consider the microscopic wetting geometry between the fiber and the droplet. Since $r<<R$, we describe the droplet as an infinite bath and the equilibrium wetting is attained in the same way as a cylinder on the surface of a liquid bath, where the liquid surface is flat and Young's law is satisfied. Overall, there is a loss of liquid-vapour and solid-vapour interface in favour of a gain of solid-liquid interface. Considering this microscopic picture, we find $\beta = \mathrm{sin}\theta_y + (\pi - \theta_y)\mathrm{cos}\theta_y $, where $\theta_y$ is the Young's angle of the liquid on the solid. 

As explained in the main manuscript, for SIS we make the qualitative observation that the bending cost of the fiber not in contact with the droplet is roughly equal in magnitude to the bending cost of the fiber being wet by the droplet. Thus, the bending cost upon winding is now twice as large $\frac{\pi E r^4 }{4R^2}$ whereas the gain in surface energy, $-2\gamma \beta r$, is unchanged. Ensuring a reduction in the total energy upon winding now yields $\alpha = \sqrt{\pi/8\beta}$. 

\section{Global Surface Energy Change Upon Winding}
\begin{figure}
     \includegraphics[width=7cm]{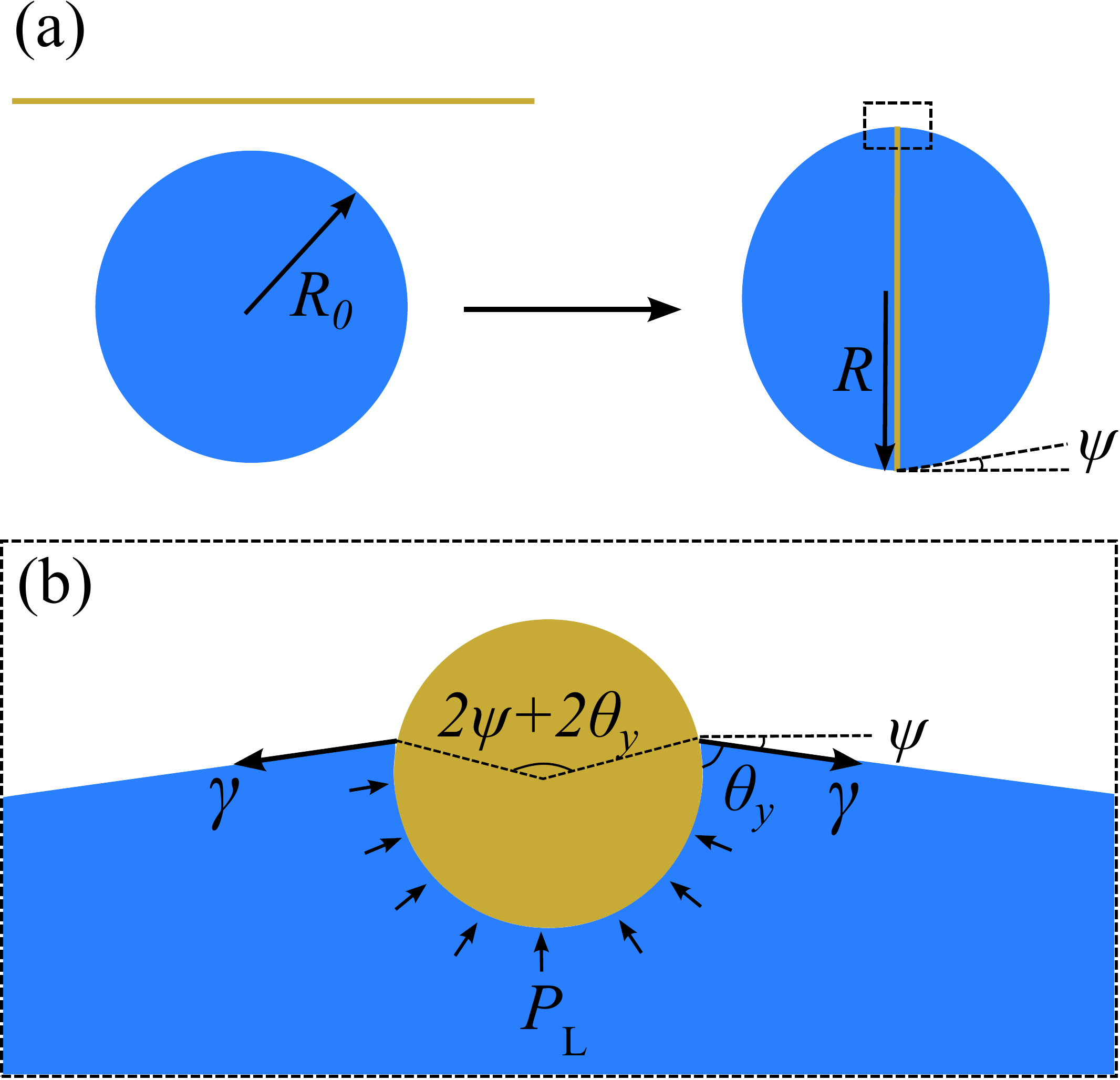}
\caption{(a) A droplet of initial size $R_0$ becomes a lens of equatorial radius $R$ upon being wound by a fiber. The lens is composed of two spherical caps which intersect the equator at an angle $\pi/2 - \psi$. (b) A zoomed-in cross-sectional view of the dashed rectangular area in (a). $P_\mathrm{L}$ denotes the Laplace pressure and $\gamma$ indicates contact line surface tension forces.}
\label{figS_lens}
\end{figure}

When considering the free-energy change upon the fiber winding the droplet, we only considered bending energy and wetting energy between the fiber and the droplet. In doing so, we ignored any global changes in area as the droplet assumes a lenticular shape. To justify this assumption, we must first examine the resultant lenticular shape which is depicted in Fig.~\ref{figS_lens} but was first discussed in ~\cite{Roman2010}. As seen in Fig.~\ref{figS_lens}(a), if we denote the radius of the initial droplet as $R_{0}$, then the radius at the equator of the lens will be denoted $R$, where in general $R > R_0$ to conserve volume. The radius of curvature of the spherical caps composing the lens will be denoted $R_\mathrm{L}$. In Fig.~\ref{figS_lens}(b), we draw the microscopic picture of the wetting between the liquid and the fiber in the wound state. The circular shape of the beam is maintained as a result of a distribution of capillary forces: contact line forces $\gamma$ and Laplace pressure $P_\mathrm{L}$. Young's angle is satisfied between the solid and the liquid.

\subsection{The Lens Configuration}
The lens configuration can be described through a force balance on the fiber. The net liquid force acting inwards per unit length is:

\begin{equation}
F_{\mathrm{net},\gamma} = 2\gamma \mathrm{sin}\psi - \frac{2\gamma }{R_\mathrm{L}}\big( 2r \mathrm{sin}(\psi + \theta _y) \big) \; ,
\end{equation}
where $\psi$ is denoted in Fig.~\ref{figS_lens}.  We can use the spherical cap identity $R/R_\mathrm{L} = \mathrm{cos}\psi$ to get:

\begin{equation}
F_{\mathrm{net},\gamma} = 2\gamma \mathrm{sin}\psi - \frac{4\gamma r}{R}\big(\mathrm{cos}\psi \, \mathrm{sin}(\psi + \theta _y) \big) \; .
\end{equation}
To maintain a circular beam, the net force per unit length acting radially inwards must be $3B_\mathrm{rod}/R^3$~\cite{Landau1986}. Thus, the lens configuration must satisfy:

\begin{equation}
\frac{3\pi Er^4}{4R^3} = 2\gamma \mathrm{sin}\psi - \frac{4\gamma r}{R}\big(\mathrm{cos}\psi \, \mathrm{sin}(\psi + \theta _y) \big) \; .
\label{force_balance}
\end{equation}

In our experiments, we observe that the lens configuration appears almost completely spherical, i.e. $\psi << 1$. Thus, to proceed further, we make the assumption $\psi << 1$, and will soon show that this is valid in our case. To first order, Eq.~(\ref{force_balance}) becomes:

\begin{equation}
\frac{3\pi Er^4}{4R^3} \approx 2\gamma \psi - \frac{4\gamma r}{R}\big(\psi \mathrm{cos} \theta _y + \mathrm{sin} \theta _y \big) \; .
\label{force_balance}
\end{equation}
We can now isolate for $\psi$ to arrive at:

\begin{equation}
\psi \approx \frac{r}{R} \frac{\frac{3\pi L_{BC}^2}{8R^2}+2 \mathrm{sin} \theta _y}{1 - \frac{2r}{R}\mathrm{cos} \theta _y} \propto \frac{r}{R}\; .
\label{psi}
\end{equation}
Therefore, we see that $\psi$ scales as $r/R$. Since $r<<R$ in our experiments and $L_{BC}/R$ is on the order of unity, $\psi << 1$ is a valid assumption. 

\subsection{Volume Conservation}
Now we consider the global change in area of a droplet becoming a lens, as depicted in Fig.~\ref{figS_lens}(a). We will limit our discussion to $\psi <<1$. To conserve volume, it follows that $R$ will only be slightly larger than $R_0$, i.e. $R = R_0 ( 1+ \delta)$, where $\delta << 1$. Thus, the statement of volume conservation from a spherical droplet to the two spherical caps composing the lens reads:

\begin{equation}
\frac{4}{3}\pi R_0^3 = \frac{2}{3} \pi \Big( \frac{R}{\mathrm{cos}\psi}\Big)^3 \big( 2- 3\mathrm{sin}\psi +\mathrm{sin}^3\psi \big) \; .
\end{equation}
If we expand the right-hand side to second-order in $\delta$ and $\psi$ we find $\delta \approx \psi/2$. 

\subsection{Change in Area}
The change in area ($\Delta \mathcal{A}$) of the droplet can be written as:

\begin{equation}
\Delta \mathcal{A} = 2\pi R^2\Big( 1+\mathrm{tan}^2\big(\frac{\pi /2 - \psi}{2} \big)  \Big) - 4\pi R_0^2 \; .
\end{equation}

We expand $\Delta\mathcal{A}$ to second-order in $\psi$ and $\delta$, because as we will see, the first-order term will vanish:

\begin{equation} 
\Delta \mathcal{A} \approx 2\pi R_0^2 (1+\delta)^2 (2-2\psi+2\psi ^2) - 4\pi R_0^2
\end{equation}
\begin{equation} 
\Delta \mathcal{A} \approx 4\pi R_0^2 (1+2\delta - \psi +\psi ^2 -2\delta \psi + \delta ^2) - 4\pi R_0^2 \; .
\end{equation}
The zeroth-order terms cancel, and inserting $\delta \approx \psi/2$, we find the first-order terms vanish as well, and we are left with:

\begin{equation}
\Delta \mathcal{A} \approx \pi R_0 ^2 \psi ^2=\pi \left(\frac{R}{1+\delta} \right)^2 \psi^2 \approx \pi R ^2 \psi ^2 \;,
\label{delta_SA}
\end{equation}
up to second-order in $\psi$. Since we know that $\psi \propto r/R$, we see that $\Delta \mathcal{A} \propto R^2  (r/R)^2 \sim r^2$. Therefore the change in surface energy from global area changes is $\Delta E_\mathcal{A} \sim \gamma r^2$. However, the change in surface energy due to wetting is $\Delta E_\mathrm{w} = -2\gamma r\beta (2\pi R) \sim rR$ for one complete wind, where $\beta$ depends on the microscopic wetting picture and is of order unity. Since $r<<R$, we see that $\Delta E_\mathrm{w} \sim rR >> \Delta E_\mathcal{A} \sim r^2$, and we can neglect any global changes in area.

\end{document}